\documentclass[english]{article}
\usepackage[T1]{fontenc}
\usepackage[latin9]{inputenc}
\usepackage{amsmath}
\usepackage{graphicx}

\makeatletter

\providecommand{\tabularnewline}{\\}

\usepackage{arxiv}
\author{
Viet Lam Phung\\
Zalo Group, VNG Corporation\\
Hanoi, Vietnam\\
$\texttt{lampv@vng.com.vn}$
\And
Huy Kinh Phan\\
Zalo Group, VNG Corporation\\
Hanoi, Vietnam\\
\And
Anh Tuan Dinh\\
Zalo Group, VNG Corporation\\
Hanoi, Vietnam\\
\And
Quoc Bao Nguyen\\
Zalo Group, VNG Corporation\\
Hanoi, Vietnam\\
Information and Communication Technology University\\
Thai Nguyen University\\
Thai Nguyen, Vietnam\\
$\texttt{baonq6@vng.com.vn}$\\
}

\makeatother

\usepackage{babel}
\begin{document}
\title{Data Processing for Optimizing Naturalness of Vietnamese Text-to-speech
System}
\maketitle
\begin{abstract}
End-to-end text-to-speech (TTS) systems has proved its great success
in the presence of a large amount of high-quality training data recorded
in anechoic room with high-quality microphone. Another approach is
to use available source of \emph{found} data like radio broadcast
news. We aim to optimize the naturalness of TTS system on the found
data using a novel data processing method. The data processing method
includes 1) utterance selection and 2) prosodic punctuation insertion
to prepare training data which can optimize the naturalness of TTS
systems. We showed that using the processing data method, an end-to-end
TTS achieved a mean opinion score (MOS) of 4.1 compared to 4.3 of
natural speech. We showed that the punctuation insertion contributed
the most to the result. To facilitate the research and development
of TTS systems, we distributed the processed data of one speaker at
https://forms.gle/6Hk5YkqgDxAaC2BU6.

\keywords{utterance selection \and prosodic punctuation \and end-to-end TTS \and speech database}
\end{abstract}

\section{Introduction}

Text-to-speech (TTS) systems play an important role in widely accepted,
interactive systems like Siri, Microsoft Cortana, and Amazon's Alexa.
However, collecting data to build those systems is costly. Typically,
a professional voice talents is recruited to read dozens of hours
of text with good coverage of target domain in an anechoic room with
high-quality microphone. The speakers should maintain constant fundamental
frequency (F0), energy, speaking rate, and articulation throughout.
There are 7000 languages in the world, and most do not receive as
much research attention as English, Spanish, Mandarin, and Japanese.
The so called low-resource language\emph{ }like Vietnamese have no
carefully recorded and annotated corpora which can be used for TTS
systems. The available speech corpora such as VOV (radio broadcast
news)~\cite{vov}, VNSpeechCorpus~\cite{mica}, VIVOS~\cite{ailab}
are small, and dedicated for automatic speech recognition (ASR). The
VAIS-1000~\cite{vais}, which is a latest database for TTS, only
consists of 1000 sentences of a speaker. Another approach is to make
use of various sources of found\emph{ }data (e.g.~radio broadcast
news, automatic speech recognition (ASR) corpora, and audiobooks)
as TTS corpora~\cite{Cooper16}. The approach became the main challenge
in the TTS evaluation of the Vietnamese Language and Speech Processing
(VLSP) 2019 ~\cite{vlsp}. In the paper, we proposed a data processing
scheme to integrate the VLSP evaluation's broadcast news corpora into
our end-to-end Vietnamese TTS system consisting of Tacotron~2~\cite{tacotron2}
and WaveGlow vocoder~\cite{glow}. Our data processing scheme consists
of two key elements: 1) utterance selection using different metrics,
and 2) prosodic punctuation insertion into text. In our experiments,
we significantly improved the naturalness of our TTS system by applying
our data processing method on training data. In the TTS evaluation
of the VLSP 2019, our system achieved a MOS of 4.1 (compared to 4.3
of natural speech); which was the best result among all participants~\cite{ZaloVLSP}. 

\section{Background}

Researchers have attempted to build high-quality Vietnamese TTS systems
in the last two decades. A text normalization method was investigated
utilizing regular expressions and language model~\cite{textnorm}.
Prosodic features such as phrase breaks proved their efficacy in improving
naturalness of Vietnamese TTS system~\cite{prosody13,prosody13b}.
Different types of acoustic models were investigated such as hidden
Markov model (HMM)~\cite{prosody13c,Glottal15}, and deep neural
network (DNN)~\cite{dnnTTS}. These HMM- and DNN-based TTS systems
are limited by the oversmoothing of generated parameters~\cite{Toda05}.
A post-filtering method was proposed to compensate for the oversmoothing
effect using non-negative matrix factorization~\cite{postfilt16}.
Recently, the use of sequence-to-sequence model in acoustic modeling~\cite{tacotron2}
in combination with neural vocoders such as WaveGlow~\cite{glow}
have enormously reduced the oversmoothing effect; thus, achieving
human quality TTS~\cite{transformertts,ZaloVLSP}. 

We need dozens of hours of training data for a speaker to build a
high-quality TTS system. One solution is to use available sources
of found data. Three types of found data: radio broadcast news, automatic
speech recognition (ASR) corpora, and audiobooks were compared; showing
that radio broadcast news is a good match as TTS corpora~\cite{Cooper16}.
Different criteria such as standard deviation of fundamental frequency
(F0), speaking rate , hypo- and hyper-articulation were explored in
utterance selection for HMM-based TTS~\cite{Cooper16} and DNN-based
TTS~\cite{oben}. We are not aware of any attempts at utterance selection
for end-to-end TTS. In the paper, we explored different metrics for
utterance selection such as misalignment errors, articulation, standard
deviation of syllable duration, non-fluency, and standard deviation
of F0. 

Traditionally, an end-to-end TTS system receives a sequence of syllables,
or words as input; thus, it has no explicit prosodic features incorporating
in the input. The prosodic features such as phrase breaks is important
for the naturalness of Vietnamese TTS system~\cite{prosody13}. In
the paper, we insert the prosodic punctuations, which corresponds
to pauses in utterance, into text. We realized that the inserted prosodic
punctuations led to stable, faster convergence of the training of
Tacotron~2. Moreover, making use of prosodic punctuation derived
from utterances is a novel way to help Tacotron~2 model learn a speaker-dependent
prosodic pattern.

\section{Data}

We used a so-called ``big training dataset'' provided by the TTS
evaluation of the VLSP 2019~\cite{vlsp}. There are 15000 utterances
of a single speaker (approximately 23 hours) with corresponding text;
which cover broadcast news. The speaker was recorded at home instead
of an anechoic room. The speaker was instructed to stay in a place
as quite as possible during each recording session. Therefore, the
data features the three types of errors 1) variation in channel conditions,
2) mismatch between text and utterance content, and 3) variation in
articulation. The variation in channel conditions are caused by microphone
conditions, recording environments, channel noise, ... As a result,
some utterances have mild background noise. The mismatch between text
and utterance content is due to misspelling, tricky text, ... When
the text was meaningless and hard to read, the speaker often gave
up and said random things. The variation in articulation features
hyper-articulation, inconsistency in speaking rate, F0. We downsampled
the speech data to 16kHz.

\section{Proposed Method}

\begin{figure}[t]
\begin{centering}
\includegraphics[width=1\columnwidth]{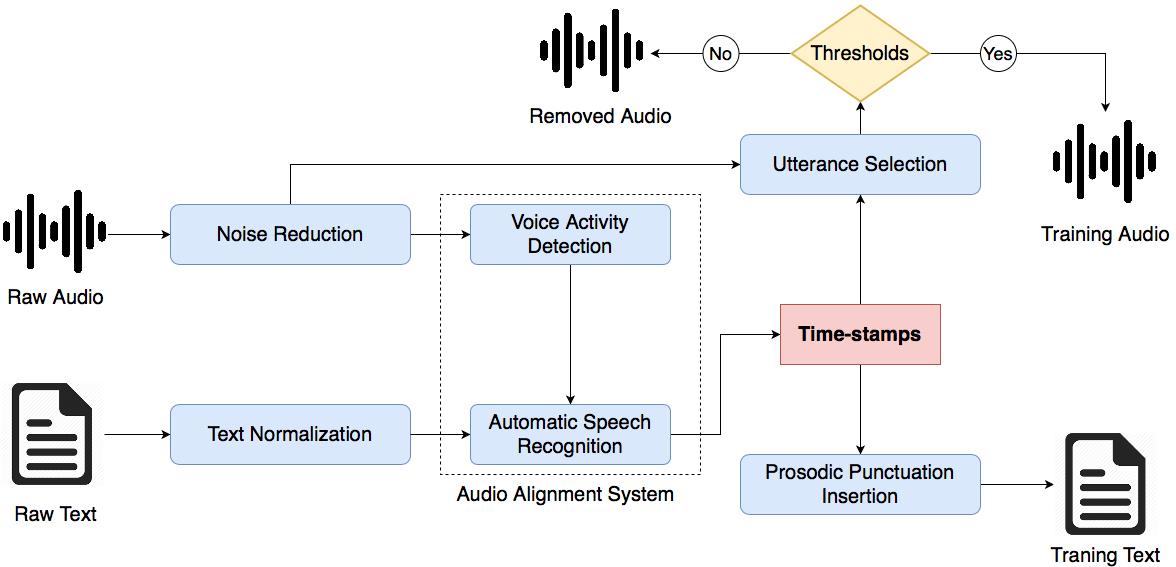}
\par\end{centering}
\caption{\label{fig:DP-scheme}Data Processing Scheme}
\end{figure}

In the section, we present our data processing scheme as shown in
Figure~\ref{fig:DP-scheme}. Given the raw utterances and corresponding
raw text, we applied a noise reduction method on the audio files using
minimum mean-squared-error estimator~\cite{Yu08}. The text was normalized
and tokenized~\cite{textnorm}. The denoised audio and corresponding
normalized text were aligned using an \emph{audio alignment system}.
Using the time-stamps obtained from the alignment, we can 1) calculate
different \emph{metrics} of utterance selection and 2) identify and
insert \emph{prosodic punctuation} to text. We selected the utterances
satisfying some experimental thresholds of the metrics. Each sentence
was spliced into phrases by prosodic punctuation; reflecting the prosody
pattern of the speaker. 

\subsection{\label{subsec:AAS}Audio Alignment System}

Vietnamese is a mono-syllabic language. We develop an audio alignment
system~\cite{Ribes97,Moreno98,Lamel02,Venkataraman04,Chan04,Boeffard12,Lanchanti15}
for Vietnamese to identify \emph{time-stamps} from audio for syllables.
We first used a voice activity detection module to segment audio into
speech and non-speech segments using adaptive context attention model~\cite{Kim18}.
It is different from~\cite{oben} where an HMM-based voice activity
detection was used. A Time Delay Neural Network (TDNN)-based ASR system~\cite{Peddinti15}
was trained over-fittingly on the speech segments and corresponding
normalized text in our data. We also biased a language model to the
normalized text~\cite{Lamel02}. Each speech segment was decoded
using the TDNN-based acoustic model and the biased language model.
The resulting time-aligned transcription was aligned with the original
normalized text; associating the obtained time-stamps to the original
text. On the other hand, matching sequences of syllables (also called
anchor points~\cite{Moreno98}) between decoding output and original
text transcript can be a good indicator that the speaker read the
sentence correctly. Moreover, we can use the time-stamps of pauses
and silences to identify prosodic punctuation.

\subsection{Utterance Selection Metrics}

We introduce different metrics addressing three type of errors in
our data:
\begin{itemize}
\item \emph{Word-error rate} (WER) of decoding output when comparing to
original text addresses the mismatch between text and utterance content.
Every utterances with a WER less than 90\% were removed from our data;
thus, we removed 800 utterances. 
\item \emph{Articulation}~\cite{Cooper16} is used to address the variation
in articulation or abnormal articulation. It is calculated as in Equation~\ref{eq:articulation}
where $P_{signal}$ is the power of speech segments extracted from
voice activity detection module; the average syllable duration (avg.syl.dur)
is calculated based on the aligned time-stamps (obtained in~\ref{subsec:AAS}).
Speech segment with high articulation is hyper articulated. The hyper-articulated
speech is unnatural because it has slow speaking rate and high energy~\cite{Hirschberg04}. 
\end{itemize}
\begin{equation}
\textrm{Articulation}=P_{signal}\times\textrm{avg.syl.dur}\label{eq:articulation}
\end{equation}

\begin{itemize}
\item \emph{Standard deviation of syllable duration }(std.syl.dur) is used
to address the inconsistency of speaking rate. The duration of each
syllable is calculated according to the aligned time-stamps. Given
a speech segment, a high value of std.syl.dur indicates that the narrator
spoke sometimes fast and sometimes slow within the segment. Speech
segments with high inconsistency of speaking rate are unnatural.
\item \emph{Non-fluency} is used to address the reading non-fluency or variation
in articulation. Moreover, the alignment procedure in~\ref{subsec:AAS}
can have misalignment errors. We can also use the non-fluency metric
to address the misalignment errors. The non-fluency is calculated
as in Equation~\ref{eq:non-fluency} where maximum duration of internal
silence and average syllable duration are calculated according to
aligned time-stamps. The internal silence is silence or pause other
than the start and end ones. A high value of non-fluency indicates
a long pause within an utterance; reflecting the non-fluency. 
\end{itemize}
\begin{equation}
\textrm{Non-fluency}=\frac{\textrm{max}\left(\textrm{internal silence duration}\right)}{\textrm{avg.syl.dur}}\label{eq:non-fluency}
\end{equation}

\begin{itemize}
\item \emph{Standard deviation of F0 }(std.F0) is used to address inconsistency
of F0. A high value of std.F0 can be due to more expressive speech.
We removed utterances with high values of std.F0.
\end{itemize}
For each metric, we rejected the 5\% of data corresponding to the
segments with the worst values of the metric.

\subsection{Prosodic Punctuation Insertion}

We detected four types of prosodic punctuation from speech based on
the duration of internal silences. The internal silence duration can
be calculated using aligned time-stamps. We represent each type of
prosodic punctuation with a special character. We then insert the
special characters into text at the positions of corresponding silences.
By experiments, we determined four ranges of silence durations to
identify the prosodic punctuations: $\left[0.12,0.15\right]$, $(0.15,0.21]$,
$(0.21,0.27]$ and more than $0.27$ second. The prosodic punctuations
are also used to mark the bad pauses caused by non-fluency. Thus,
we can prevent the models to align the bad pause frames to any syllables.

\section{Vietnamese TTS system}

Our end-to-end TTS system have two components: 1) a encoder-decoder
acoustic model and 2) a neural vocoder. The encoder-decoder acoustic
model converts a sequence of syllables with prosodic punctuations
to a 80-dimensional Mel-spectrogram. The neural vocoder generates
speech from the 80-dimensional Mel-spectrogram. In normalized text,
we consider tokens, inserted prosodic punctuation as syllables. In
the paper, we utilized Tacotron~2~\cite{tacotron2} for acoustic
modeling, and WaveGlow vocoder~\cite{glow}. As a result, our end-to-end
system can achievee a real-time inference speed.

\subsection{Encoder-decoder acoustic model}

Generally, acoustic model in neural TTS system has an attention-based
encoder-decoder structure. Given encoder outputs $\textbf{x}=[\textbf{x}_{1,}\textbf{x}_{2},...,\textbf{x}_{n}]$
as memory entries and previous decoder hidden state $\textbf{h}_{i-1}$,
for each decoder output step $i$, an energy value $e_{i,j}$ is calculated
for each $\textbf{x}_{j}$ by a trainable attention mechanism~\cite{tacotron,tacotron2}
as in Equation~\ref{eq:att1}. The energy is normalized to obtain
alignment vector $\alpha_{i}$ as in Equation~\ref{eq:att2}; then,
we produce the context vector $\textbf{c}_{i}$ from the alignment
vector $\alpha_{i}$ as in Equation~\ref{eq:att3}. 
\begin{equation}
e_{i,j}=\textrm{Attention}\left(\textbf{h}_{i-1},\textbf{x}_{j}\right)\label{eq:att1}
\end{equation}

\begin{equation}
\alpha_{i,j}=\frac{\exp\left(e_{i,j}\right)}{\sum_{k=1}^{n}\exp\left(e_{i,k}\right)}=\textrm{softmax}\left(e_{i,:}\right)_{j}\label{eq:att2}
\end{equation}

\begin{equation}
\textbf{c}_{i}=\sum_{j=1}^{n}\alpha_{i,j}\textbf{x}_{j}\label{eq:att3}
\end{equation}

\subsection{WaveGlow vocoder}

In the paper, we used WaveGlow~\cite{glow} as for neural vocoding.
WaveGlow is a deep generative model for waveform generation that incorporates
Glow, a generative model for image processing,~\cite{glow_b} with
WaveNet~\cite{wavenet}. During training, a speech waveform $\textbf{y}$is
converted to a Gaussian white noise $\textbf{z}$. Conversely, a Gaussian
white noise is converted to a speech waveform by the inverse operation
during inference process. By introducing the invertible 1 $\times$
1 convolution and affine coupling layers, the loss function of the
WaveGlow vocoder is calculated as in Equation~\ref{eq:wg}. 
\begin{equation}
-\log p_{\textrm{\ensuremath{\theta}}}\left(\textbf{x}\right)=\frac{\textbf{z}(\textbf{x}^{\textrm{T}})\textbf{z}(\textbf{x})}{2\sigma_{\textrm{WG}}^{2}}-\sum_{j=0}\log\textbf{s}_{j}\left(\textbf{x},\textbf{f}\right)-\sum_{k=0}\log\lvert\textrm{det}(\textbf{W}_{k})\lvert\label{eq:wg}
\end{equation}

The $\textbf{\ensuremath{\theta}}$ denotes network parameters; $\textbf{f}$
is conditional acoustic features. The $\textbf{s}_{j}$, $\textbf{W}_{k}$,
and $\sigma_{\textrm{WG}}^{2}$ are output coefficients of $j$th
WaveNet in the affine coupling layers, the $k$th weighting matrix
of the invertible 1 $\times$ 1 convolution layer, and the assumed
variance of the Gaussian distribution, respectively.

\begin{figure}
\begin{centering}
\includegraphics[width=0.8\columnwidth]{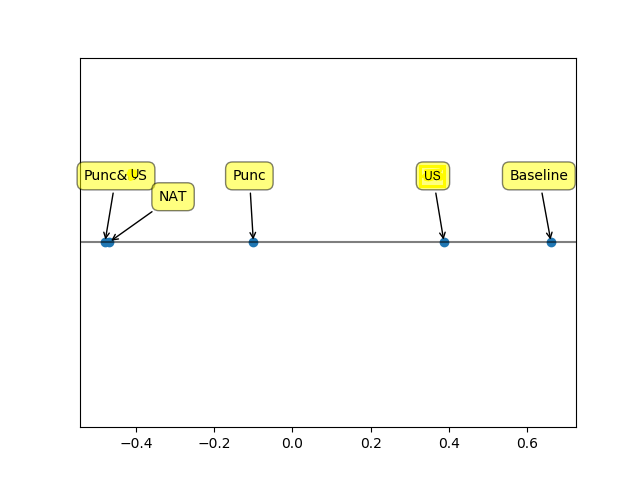}
\par\end{centering}
\caption{\label{fig:MDS}Multiple dimensional scaling results. Closer to NAT
is better}
\end{figure}

\section{Experiment}

In the section, we trained a end-to-end TTS system on provided original
data (or ``big training dataset'') as our \emph{Baseline} system.
We then evaluated the efficacy of our data processing scheme by comparing
systems trained on the processed data to the baseline system. The
\emph{US} denotes that the original data was processed by utterance
selection alone. The \emph{Punc} denotes that only prosodic punctuation
was used on original data. The \emph{Punc\&US} denotes that both utterance
selection and prosodic punctuation insertion were used. In total,
we trained four end-to-end TTS systems. We leave-out 32 sentences
for testing. With each case of data processing, we used 90\% or remaining
sentences for training and the other 10\% validation. Only training
and validation text has prosodic punctuations but not testing text.
Our training configurations were the same as in~\cite{tacotron2,glow}.
We submitted the Punc\&US system to the VLSP 2019's evaluation. The
Nat denotes target natural speech.

\begin{table}
\begin{centering}
\begin{tabular}{|c|c|c|c|c|}
\hline 
A \textbackslash{} B & Punc & US & Punc\&US & NAT\tabularnewline
\hline 
\hline 
Baseline & $-$0.81{*} & $-$0.31{*} & $-$1.03{*} & $-$1.16{*}\tabularnewline
\hline 
Punc &  & 0.72{*} & $-$0.28{*} & $-$0.75{*}\tabularnewline
\cline{1-1} \cline{3-5} 
US & \multicolumn{1}{c}{} &  & $-$0.56{*} & $-$0.97{*}\tabularnewline
\cline{1-1} \cline{4-5} 
Punc\&US & \multicolumn{1}{c}{} & \multicolumn{1}{c}{} &  & $-$0.53{*}\tabularnewline
\hline 
\end{tabular}
\par\end{centering}
\caption{\label{tab:CMOS}Comparative MOS results. Positive values indicate
A is better than B. Results marked with an asterisk are significantly
different ($p$ < 0.05) as compared to $0$ (representing no preference)
in a $1$-sample $t$-test.}
\end{table}

\subsection{VLSP 2019's Evaluation}

A MOS test was conducted to compare our submitted system to other
participants in the evaluation. There are more than 30 participating
groups from academy and industry in the evaluation. There are 20 test
sentences issued by the organizer of VLSP. We did not insert prosodic
punctuations into test sentence because we do not have audio files
to detect the punctuation. There are 24 testers. At each trial, a
listener was asked to rate the quality of a utterance in a 5-point
scale: ``excellent'' (5), ``good'' (4), ``fair'' (3), ``poor''
(2), ``bad'' (1). Our system achieved a MOS result of 4.1 (compared
to 4.3 of target NAT)~\cite{ZaloVLSP}. The result suggested that
using our data processing method is efficient in optimize the naturalness
of end-to-end TTS system. Moreover, the MOS result is the best result
among the 30 participating groups in the evaluation.

\subsection{Comparative Evaluation}

We conducted a comparative MOS (CMOS) test to explore the contribution
of the proposed method to the naturalness of end-to-end TTS systems.
At each trial, participants listen to samples A and B in sequence
and were then asked: ``Is A more natural than B?'' Responses were
selected from a 5-point scale that consisted of ``definitely better''
($+$2), ``better'' ($+$1), ``same better'' (0), ``worse''
($-$1), ``definitely worse'' ($-$1). The test involved 32 sentences,
and 10 system pairs; resulting in 32 $\times$ 10 = 320 unique trials.
We limited each listener to hear each unique sentence once (presentation
order was randomized); therefore we need 320 $\div$ 32 = 10 listener
to cover all trials. We recruited 20 participants who are native Vietnamese
speakers. Table~\ref{tab:CMOS} shows the pair-wise relative quality
of the systems. 

To approximate the ordering between all systems, we projected the
non-negative pair-wise relative quality matrix to a single dimension
using multiple dimensional scaling (MDS). Figure~\ref{fig:MDS} shows
the results. All data processing methods can improve the quality of
TTS system. The results suggested that using our proposed method is
efficient in optimizing the naturalness of end-to-end TTS system.
By using our data processing method (Punc\&US), which includes both
utterance selection and punctuation insertion, we achieved close quality
to natural speech (NAT). Interestingly, the prosodic punctuation insertion
(Punc) is more efficient than utterance selection (US) in improving
the quality of TTS system. 

\section{Conclusion}

In this paper, we proposed a data processing technique including utterance
selection and prosodic punctuation insertion. We showed that using
the data processing method can improve the quality of end-to-end TTS
system trained on found data. In a VLSP 2019's evaluation, our system
trained on processed data achieved a MOS result of 4.1; which the
best MOS result among participants in the evaluation. Our CMOS test
showed that the punctuation insertion contributed more to the result
than the utterance selection. All of the processing methods can improve
the quality of end-to-end TTS system trained on found data. In future
works, we will predict the prosodic punctuation for test sentence
from text. We distributed the processed data of one speaker at https://forms.gle/6Hk5YkqgDxAaC2BU6.

\bibliographystyle{IEEEtran}
\bibliography{references}

\end{document}